\documentclass{article}
              \usepackage{romp33}

\font\fr=eufm10  scaled \magstep 1   
\font\ddpp=msbm9  scaled \magstep 1  
\font\ddppp=msbm6  scaled \magstep 1  

\def\QED{\hskip0.1em\hfill\null\ \null\nobreak\hfill
\kern3pt\lower1.8pt\vbox{\hrule\hbox   {\vrule\kern1pt\vbox{\kern1.7pt
\hbox{$\scriptstyle   QED$}\kern0.2pt}\kern1pt\vrule}\hrule}}

\def\R{\hbox{\ddpp R}}               
\def\Rp{\hbox{\ddppp R}}               

\def\hfl#1#2{\smash{\mathop{\hbox to 12 mm{\rightarrowfill}}
\limits^{\scriptstyle#1}_{\scriptstyle#2}}}

\begin{document}

 \title{\bf Variational integrators and time-dependent lagrangian systems}

\author{M. DE LE\'ON, D. MART\'IN DE DIEGO\thanks{ \hspace{0.2cm}Research in part supported by grants DGICYT (Spain), PB97-1257
and PGC2000-2191-E.}
\\ Instituto de Matem\'aticas y F\'\i sica Fundamental,
CSIC,\\
Serrano 123, 
28006 Madrid, Spain
 \\ (e-mail: mdeleon{@}imaff.cfmac.csic.es, d.martin{@}imaff.cfmac.csic.es)} 
\date{\today}          
\maketitle

\begin{abstract}
This paper presents a method to construct  variational integrators for time-dependent lagrangian systems. The resulting algorithms are symplectic, preserve the momentum map associated with a Lie group of symmetries and also describe the energy variation. 
\end{abstract}


\section{Introduction}

There are several numerical integration methods \cite{SaCa} that preserve some of the invariants of an
autonomous mechanical system.  In  \cite{Lee1}, T.D. Lee studies the possibility that time can be regarded as a 
{\it bona fide} dynamical variable  giving a discrete time formulation of mechanics (see also \cite{Lee2,Lee3}).  From other point of view (integrability aspects)  Veselov \cite{veselov1} uses a discretization of the equations of classical mechanics. Both approaches can be characterized as the creation of integrators based on a discretization of the variational principle determined by a lagrangian function. These integration methods have usually better long term simulation properties and computational efficiency than the conventional ones. 

The main geometrical invariants that these integrators preserve are symplecticity, energy or/and  momentum. Ge and Marsden \cite{Ge-Marsden}  have proved that a  constant time stepping integrator cannot preserve  the symplectic form, energy and momentum, simultaneously, unless it coincides with the exact solution of the initial system up to a time reparametrization.
However,  Kane, Marsden, Ortiz and  West \cite{Kane} show that using an appropriate definition of symplecticity and an adaptative time stepping it is possible to construct a variational integrator which is simultaneously symplectic, momentum and energy preserving.

The purpose of this paper is to extend the results previously obtained for conservative mechanical systems to the case of time-dependent lagrangian systems following Lee's approach. The method is also based on a  discrete variational principle. 
In a further work, we will test our algorithm in relevant  
examples and analyze the deep relationship between the continuous and discrete cases.

From the point of view of applications to problems in mechanics  
(time-dependent harmonic oscillator, Meshchersky's equations in rocketry, for instance), and 
in control theory, the time-dependent case definitely deserves special attention. 
Moreover, time-dependent lagrangian systems appear as an indispensable tool in many economic 
problems~\cite{Ra,SaRa}. In fact, a typical optimization problem in modern economics deals with extremizing 
the functional 
$
\int^T_0 D(t)U[f(t,k,\dot{k})]\; dt 
$
subject or not to constraints. Here, $D(t)$ is a discount rate factor, $U$ an utility function, 
$f$ a consumption function and $k$ the capital-labor ratio.
Moreover, the time-dependent case is also important because its relation with Classical Field Theory,
and in a future work, we  will investigate the extension of these ideas in the
framework of multisymplectic geometry 
(see the preliminary work by Marsden, Patrick and  Shkoller \cite{Ma2},
and \cite{CoMa2,LMD} for variational systems subjected to constraints 
and nonholonomic mechanical systems).

\section{Time-dependent Lagrangian systems}

Time-dependent lagrangian theory is sometime studied using
an homogeneous formalism. If $L: \R\times TQ\rightarrow \R$ is a time-dependent 
lagrangian then the corresponding homogeneous lagrangian $\bar{L}: T_0(\R\times Q)\rightarrow \R$   
is defined  by:
\[
\bar{L}(t, q , \dot{t}, \dot{q})=\dot{t} L(t, q, \dot{q}/\dot{t})
\]
(we delete the submanifold $\dot{t}=0$ from the whole tangent space $T(\R\times Q)$.) 
A solution $(t(s), q(s))$ of the Euler-Lagrange equation for $\bar{L}$ verifies:
\begin{eqnarray*}
&&\frac{d}{ds}\left(\frac{\partial {L}}{\partial \dot{q}}(t(s), q(s), \frac{\dot{q}(s)}{\dot{t}(s)})\right)-\dot{t}(s)\frac{\partial {L}}{\partial q}(t(s), q(s), \frac{\dot{q}(s)}{\dot{t}(s)})=0\\
&&\frac{d}{ds}\left({L}(t(s), q(s), \frac{\dot{q}(s)}{\dot{t}(s)})-\frac{\dot{q}(s)}{\dot{t}(s)}\frac{\partial {L}}{\partial \dot{q}}(t(s), q(s), \frac{\dot{q}(s)}{\dot{t}(s)})\right)-\dot{t}(s)\frac{\partial {L}}{\partial t}(t(s), q(s), \frac{\dot{q}(s)}{\dot{t}(s)})=0
\end{eqnarray*}

Along the submanifold $\dot{t}=1$ these equations represent the Euler-Lagrange equations for 
$L$ and the non-preservation of the energy function $E_L=\dot{q}\partial L/\partial \dot{q}-L$, 
respectively (see \cite{CCCI}, for instance).

\section{The time-dependent variational integrator}\label{time}

A discrete time dependent lagrangian 
is a map $L_d: \R\times Q\times \R\times Q \rightarrow \R$.
Define the action sum ${\cal S}: (\R\times Q)^{N+1}\rightarrow \R$ corresponding to the lagrangian $L_d$ by
\[
{\cal S}=\sum_{k=0}^{N-1} (t_{k+1}-t_k) L_d(t_k, q_{k}, t_{k+1}, q_{k+1}),
\]
where $q_k\in Q$ and  $t_k\in\R$ for $0\leq k\leq N$. 

We know that for any product manifold $M_1\times \ldots\times  M_a$,
$T_{(x_1, \ldots, x_a)}^*(M_1\times\ldots\times  M_a)\simeq T^*_{x_1} M_1\oplus\ldots\oplus  T^*_{x_a} M_a$, 
for all $x_i\in M_i$, $1\leq i\leq a$.
Therefore, any covector $\alpha\in T_{(x_1,\ldots,  x_a)}^*(M_1\times\ldots\times  M_a)$
admits a decomposition $\alpha=\alpha_1+\ldots +\alpha_a$ where $\alpha_i\in T^*_{x_i} M_i$, $1\leq i\leq a$.
Thus,
\begin{eqnarray*}
dL_d(t_0, q_0, t_1, q_1)&=&D_{1} L_d(t_0, q_0, t_1, q_1)+D_{2} L_d(t_0, q_0, t_1, q_1)\\
&&+D_{3} L_d(t_0, q_0, t_1, q_1)+D_{4} L_d(t_0, q_0, t_1, q_1)
\end{eqnarray*}
since 
$
T^*_{(t_0, q_0, t_1, q_1)}(\R\times Q\times \R\times Q)=T^*_{t_0} 
\R\oplus T^*_{q_0}Q\oplus T^*_{t_1}\R\oplus T^*_{q_1}Q
$
and, in addition, we have
\[
dL_d(t_0, q_0, t_1, q_1)=\bar{D}_{1} L_d(t_0, q_0, t_1, q_1)+\bar{D}_{2} L_d(t_0, q_0, t_1, q_1)
\]
since
$
T^*_{(t_0, q_0, t_1, q_1)}(\R\times Q\times \R\times Q)=T^*_{(t_0, q_0)} (\R\times Q)
\oplus  T^*_{(t_1, q_1)}(\R\times Q)\; .
$
From now on, we will use both decompositions ($D$ and $\bar{D}$) of 1-forms on $(\R\times Q)^2$.

The discrete variational principle assures that the solutions 
of the discrete system must extremize the action sum given fixed points $(t_0, q_0)$ and $(t_N, q_N)$.
Extremizing ${\cal S}$ over $t_k$, $q_k$, $1\leq k\leq N-1$ leads to the following system of equations:
\begin{equation}\label{asd}
(t_{k+1}-t_k)D_2L_d(t_k, q_k,t_{k+1}, q_{k+1})+(t_k-t_{k-1})D_4L_d(t_{k-1}, q_{k-1}, t_k, q_{k})=0\\
\end{equation}

\begin{minipage}{11cm}
\begin{eqnarray*}
&&(t_{k+1}-t_k)D_1L_d(t_k, q_k, t_{k+1}, q_{k+1})-L_d(t_k, q_{k}, t_{k+1}, q_{k+1})dt_k+\\
&&\qquad +(t_{k}-t_{k-1})D_3L_d(t_{k-1}, q_{k-1}, t_k,  q_{k})+L_d(t_{k-1}, q_{k-1}, t_k, q_{k})dt_k=0\; ,
\end{eqnarray*}
\end{minipage}\hfill
\begin{minipage}{1cm}
\begin{equation}
\label{asd3}
\end{equation}
\end{minipage}

\

or, equivalently,
\begin{equation}\label{asd1}
\bar{D}_1\left[(t_{k+1}-t_k)L_d(t_k, q_k,t_{k+1}, q_{k+1})\right]+
\bar{D}_2\left[ (t_{k}-t_{k-1})L_d(t_{k-1}, q_{k-1}, t_k, q_{k})\right]=0\; .
\end{equation}

Take $\bar{L}_d(t_k, q_k, h_k, q_{k+1})=L_d(t_k, q_k,  t_k+h_k, q_{k+1})$, 
where $h_k=t_{k+1}-t_k$ is the time step, 
then
\begin{eqnarray*}
\frac{\partial L_d}{\partial t_k}&=&\frac{\partial  \bar{L}_d}{\partial t_k}-\frac{\partial \bar{L}_d}{\partial h_k}\; ,\\
\frac{\partial L_d}{\partial t_{k+1}}&=& \frac{\partial\bar{L}_d}{\partial h_k}\; .\\
\end{eqnarray*}
Therefore the second set of equations is equivalent to:

\parbox{10cm}{
\begin{eqnarray*}
&&h_k \frac{\partial \bar{L}_d}{\partial t_k} (t_k, q_k, h_k, q_{k+1})-h_k 
\frac{\partial \bar{L}_d}{\partial h_k}(t_k, q_k, h_k, q_{k+1})-
\bar{L}_d (t_k, q_k, h_k, q_{k+1})+\\
&&+ h_{k-1} \frac{\partial \bar{L}_d}{\partial h_{k-1}} (t_{k-1}, q_{k-1}, 
h_{k-1}, q_k)+\bar{L}_d (t_{k-1}, q_{k-1}, h_{k-1}, q_k)=0\; .
\end{eqnarray*}}
\hfill\parbox{1cm}{
\begin{equation}\label{nueva}
\end{equation}
}

Next, we define   the discrete energy function  as (see Kane {\em et al} \cite{Kane}):
\begin{eqnarray*}
E_d (t, x, h, y)&=&-\frac{\partial  }{\partial h}\left[h \bar{L}_d(t, x, h, y)\right]= 
-\bar{L}_d (t, x, h, y)-hD_3 \bar{L}_d(t, x, h, y)\; .\\
\end{eqnarray*}
One can also motivate this definition using  the discrete variational principle (see subsection \ref{www}).

We deduce that  equation (\ref{nueva}) is equivalent to:
\[
\frac{1}{h_k}\left[E_d(t_k, q_k, h_k, q_{k+1})-E_d(t_{k-1}, q_{k-1}, h_{k-1}, q_{k})\right]=
-\frac{\partial  \bar{L}_d}{\partial t_k} (t_k, q_k, h_{k}, q_{k+1})
\]
which  is a ``discretization'' of the well-known equation
${dE_L}/{dt}=-{\partial L}/{\partial t}
$
for continuous time-dependent lagrangian systems, that measures the non-conservation of the energy. 

Consider the mapping
\[
\Phi: (\R\times Q)^2\longrightarrow  (\R\times Q)^2
\]
defined by
\[
\Phi(t_{k-1}, q_{k-1}, t_k, q_{k})=(t_{k}, q_{k}, t_{k+1}, q_{k+1}), \  1\leq k\leq N-1\; ,
\]
where $q_{k+1}$, $t_{k+1}$ are implicitly defined by Equations (\ref{asd}). 
$\Phi$ will be called the {\em discrete flow} of the discrete lagrangian $L_d$.
If the matrix
$
\left(
\bar{D}_{12}L_d
\right)
$
is regular, then the discrete map $\Phi$ exists and  is uniquely defined .

\begin{example}{Example}{\bf A single particle in one-dimensional space}
{\rm 

Take the continuous  lagrangian
$\displaystyle{
L(t, x, v)=\frac{m}{2}v^2-V(t,x)
}$
and a typical discretization
\[
L_d(q_0, t_0, q_1, t_1)=\frac{m}{2}\left(\frac{q_1-q_0}{t_1-t_0}\right)^2-
V\left(\frac{q_1+q_0}{2},\frac{t_1+t_0}{2}\right)
\]
The implicit discrete algorithm is given by:
\begin{eqnarray*}
&&m\left( \frac{q_{k+1}-q_k}{t_{k+1}-t_k}-\frac{q_k-q_{k-1}}{t_{k}-t_{k-1}}\right)\\
&&=
-\frac{1}{2}\left[(t_{k+1}-t_k)\frac{\partial V}{\partial x}\left(\frac{q_{k+1}+q_k}{2}, 
\frac{t_{k+1}+t_k}{2}\right)+(t_k-t_{k-1})\frac{\partial V}{\partial x}\left(\frac{q_k+q_{k-1}}{2} ,
\frac{t_k+t_{k-1}}{2}\right)\right]\; ,\\
&&E_L(\frac{t_{k+1}+t_k}{2}, \frac{q_{k+1}+q_k}{2}, \frac{q_{k+1}-q_k}{t_{k+1}-t_k})-
E_L(\frac{t_k+t_{k-1}}{2}, \frac{q_k+q_{k-1}}{2}, \frac{q_k-q_{k-1}}{t_k-t_{k-1}})\\
&&=
\frac{1}{2}\left[(t_{k+1}-t_k)\frac{\partial V}{\partial t}\left(\frac{q_{k+1}+q_k}{2}, 
\frac{t_{k+1}+t_k}{2}\right)+(t_k-t_{k-1})\frac{\partial V}{\partial t}\left(\frac{q_k+q_{k-1}}{2}, 
\frac{t_k+t_{k-1}}{2}\right)\right]
\end{eqnarray*}
where $E_L(t, x, v)=\frac{m}{2}v^2+V(t,x)$ is the energy function.
}
\end{example}

\subsection{Convergence and error analysis for the discrete algorithm}

For simplicity, we suppose that $Q$ is a vector space and choose the 
following discretization of the lagrangian $L: \R\times TQ\rightarrow \R$:
\[
{L}_d(t_k, q_k, t_{k+1}, q_{k+1})=L(\frac{t_k+t_{k+1}}{2}, \frac{q_k+q_{k+1}}{2}, \frac{q_{k+1}-q_k}{t_{k+1}-t_k})
\]
In terms of the discrete velocity $v_k=(q_{k+1}-q_k)/(t_{k+1}-t_k)$ and the time step $h_k$,  
\[
\tilde{L}_d(t_k, q_k, h_{k}, v_{k})=L(t_k+\frac{h_k}{2}, q_k+v_k\frac{h_k}{2}, v_k)
\]
Then, equations (\ref{asd}) and (\ref{asd3}) become
\begin{eqnarray*}
&&\frac{1}{2h_k}\left[h_k\frac{\partial \tilde{L}_d}{\partial q_k}(V_k)+h_{k-1}\frac{\partial \tilde{L}_d}{\partial q_{k-1}}(V_{k-1})\right]-\frac{1}{h_k}\left[
\frac{\partial \tilde{L}_d}{\partial v_k}(V_k)-\frac{\partial \tilde{L}_d}{\partial v_{k-1}}(V_{k-1})\right]=0\\
&&\frac{1}{h_k}\left[
v_k\frac{\partial \tilde{L}_d}{\partial v_k}(V_{k})-\tilde{L}_d(V_{k})
-v_{k-1}\frac{\partial \tilde{L}_d}{\partial v_{k-1}}(V_{k-1})-\tilde{L}_d(V_{k-1})\right]\\ 
&&=-\frac{1}{2h_k}\left[h_k \frac{\partial \tilde{L}_d}{\partial t_k}(V_k)
+h_{k-1} \frac{\partial \tilde{L}_d}{\partial t_{k-1}}(V_{k-1})\right]
\end{eqnarray*}
where $V_i=(t_i, q_i, h_i, v_i)$.
It is clear that if $h_{k-1}\rightarrow 0$ and $h_{k}\rightarrow 0$ 
then the previous equations converge respectively to:
\[
\frac{\partial L}{\partial q}-\frac{d}{dt}\left(\frac{\partial L}{\partial \dot{q}}
\right)=0\qquad \hbox{and}\qquad \frac{d}{dt}E_L=-\frac{\partial L}{\partial t}
\]
As an  alternative way, we consider the  discretization of the homogeneous  lagrangian:
\[
{\cal L}_d(t_k, q_k, t_{k+1}, q_{k+1})=\frac{t_{k+1}-t_k}{\bar{h}}L(\frac{t_k+t_{k+1}}{2}, \frac{q_k+q_{k+1}}{2}, \frac{q_{k+1}-q_k}{t_{k+1}-t_k})
\]
where $\bar{h}\in \R_+$ is the time step. 
We then extremize
\[
{\sc S}=\sum_{k=0}^{N-1}{\cal L}_d(t_k, q_k, t_{k+1}, q_{k+1})=\frac{1}{\bar{h}}\sum_{k=0}^{N-1} (t_{k+1}-t_k)L(\frac{t_k+t_{k+1}}{2}, \frac{q_k+q_{k+1}}{2}, \frac{q_{k+1}-q_k}{t_{k+1}-t_k})
\]
to derive equations (\ref{asd}) and (\ref{asd3}) for the function 
\[
L_d(t_k, q_k, t_{k+1}, q_{k+1})=L(\frac{t_k+t_{k+1}}{2}, \frac{q_k+q_{k+1}}{2}, \frac{q_{k+1}-q_k}{t_{k+1}-t_k})
\]

The above formalism  is useful since now we can use the results of Chapter 
6 of \cite{GL} to analyze the local truncation and global error  
for (\ref{asd}) and (\ref{asd3}) as a particular case.

\subsection{Symplecticity of the algorithm}\label{www}

Now, we will show that the algorithm (\ref{asd}) is symplectic in a natural way. 
By similar arguments to those used in \cite{Kane}, 
it is easy to prove that for solutions of the algorithm (\ref{asd}) without imposing endpoints conditions: 
\begin{eqnarray*}
d{\cal S}_{|solutions}&=& (t_1-t_0) D_2 L_d(t_0, q_0, t_1, q_1) +(t_N-t_{N-1}) D_4 L_d( t_{N-1}, q_{N-1},
  t_N, q_N)\\
&&+(t_1-t_0) D_1 L_d (t_0, q_0, t_1, q_1) - L_d (t_0, q_0,  t_1, q_1)dt_0\\
&&+ (t_N-t_{N-1}) D_3 L_d (t_{N-1}, q_{N-1}, t_N, q_N)+L_d (t_{N-1}, q_{N-1}, t_N, q_N)dt_N\\
&=&
(t_1-t_0) D_2 L_d(t_0, q_0, t_1, q_1) +(t_N-t_{N-1}) D_4 L_d( t_{N-1}, q_{N-1},
  t_N, q_N)\\
&&+h_0D_1 \bar{L}_d (t_0, q_0, h_0, q_1) + E_d(t_0, q_0, h_0, q_1) dt_0\\
&&-E_d (t_{N-1}, q_{N-1}, h_{N-1}, q_{N})d t_N\; .\\
\end{eqnarray*}
where $h_k=t_{k+1}-t_k$.
If we put
\begin{eqnarray*}
\hbox{Energy}_{t_0}&=& h_0\frac{\partial  \bar{L}_d}{\partial t_0}  (t_0, q_0, h_0, q_1) + E_d(t_0, q_0, h_0, q_1)\\
\hbox{Energy}_{t_N}&=&E_d (t_{N-1}, q_{N-1}, h_{N-1},  v_{N})
\end{eqnarray*}
then
\begin{eqnarray*}
dS_{|{solutions}}&=&h_0 D_2 \bar{L}_d(t_0, q_0, t_1, q_1)
 +h_{N-1} D_4 \bar{L}_d( t_{N-1}, q_{N-1}, h_{N-1},  q_N)\\
&&+\hbox{Energy}_{t_0}dt_0-\hbox{Energy}_{t_N}d t_N\; .
\end{eqnarray*}
Observe the similarity of this expression  with the classical transversality 
conditions for variable initial and terminal points when we are studying extremals 
of functionals determined by time-dependent lagrangian systems. 

Now, fix $N=2$ and consider the 1-forms $\Theta_{L_d}^-$ and 
$\Theta_{L_d}^+$ following a similar definition than in Marsden {\em et al} \cite{Ma2}:
\begin{eqnarray*}
\Theta_{L_d}^-(t_0, q_0, t_1, q_1)&=&(t_1-t_0) D_2 L_d(t_0, q_0, t_1, q_1)\\
&&+ (t_1-t_0) D_1L_d (t_0, q_0,  t_1, q_1)-L_d (t_0, q_0, t_1, q_1)\, dt_0\\
\Theta_{L_d}^+(t_0,q_0, t_1,  q_1)&=&(t_1-t_0) D_4 L_d(t_0, q_0, t_1, q_1)\\
&&+(t_1-t_0) D_3L_d (t_0, q_0, t_1, q_1)+L_d (t_0, q_0, t_1, q_1)\, dt_1\; .\\
\end{eqnarray*}

In view  of the equations defining the algorithm, we deduce that 
\[
d {\cal S}_{|solutions}=\Theta_{L_d}^- + \Phi^*\Theta_{L_d}^+
\]
Since $dd{\cal S}=0$, we have  
$
d\Theta_{L_d}^-=-\Phi^*d\Theta_{L_d}^+ ,
$
but observe that
$
\Theta_{L_d}^-+\Theta_{L_d}^+=d((t_1-t_0)L_d),
$
that is, $d\Theta_{L_d}^- + d\Theta_{L_d}^+=0$
and, consequently,
$
\Phi^*\Omega_d=\Omega_d
$
where $\Omega_d=d\Theta_{L_d}^-=-d\Theta_{L_d}^+$.

Notice that $\Omega_d=-d\Theta_{L_d}^+$, where $\Theta_{L_d}^+$ is equal to
\[
h_0 D_4L_d (t_0, q_0, t_1, q_1)-E_d (t_0, q_0, h_0, q_1)dt_1
\]
with $t_0+h_0=t_1$.

The 2-form $\Omega_d$ is symplectic if and only if the matrix $(\bar{D}_{12}L_d)$ 
is regular. Observe also the similarity of the constructions of  $\Theta_{L_d}^+$ and 
$\Omega_d$ with the Poincar\'e-Cartan 1-form and 2-form, respectively, 
for time-dependent lagrangian systems (see \cite{LR,Saunders})

\subsection{Momentum conservation}

The conservation of momentum for an algorithm invariant under a symmetry group follows 
in the same way as usual \cite{WeMa}  considering here the invariance of the function 
$(t_{k+1}-t_k)L_d(t_k, q_k, t_{k+1}, q_{k+1})$.  

In particular, we suppose that $(t_{k+1}-t_k)L_d(t_k, q_k, t_{k+1}, q_{k+1})$ is 
$G$-invariant under the diagonal action of a Lie group $G$ on $\R\times Q$. 
Denote by $\Phi_g : \R\times Q \longrightarrow \R\times Q$  the group action, $g \in G$. 
For any element $\xi$ in the Lie algebra
$\hbox{\fr g}$ of $G$, we denote by
$\xi_{\Rp\times Q}$ the fundamental vector field which generates the flow
$\Phi_{\exp (s\xi)}: \R\times Q\rightarrow \R\times Q$. 
Moreover $\xi_{\Rp\times Q}(t,q)=\xi_{\Rp}(t,q)+\xi_Q(t,q)$ where $\xi_{\Rp}(t,q)
\in T_{t}\R$ and $\xi_{Q}(t,q)\in T_q Q$.

Therefore, applying the results for discrete algorithms for 
conservative mechanical systems (see for instance \cite{WeMa}) 
we obtain that the momentum map $J_d: (\R\times Q)^2\rightarrow \hbox{\fr g}^*$ defined by
\[
\langle J_d(t_k, q_k, t_{k+1}, q_{k+1}), \xi\rangle = \langle \bar{D}_2\left[(t_{k+1}-t_k)L_d(t_k, q_k, t_{k+1}, q_{k+1})\right], \xi_{\Rp\times Q}(t_{k+1}, q_{k+1})\rangle
\]
is preserved by the discrete flow $\Phi: (\R\times Q)^2\longrightarrow  (\R\times Q)^2$.

\section{Symplectic-Energy-Momentum preserving variational integrators}

In Kane, Marsden and Ortiz \cite{Kane} it is constructed a variational 
integrator for conservative mechanical systems with time-step adaptation. 
Their algorithm is symplectic, energy and momentum conserving. We will show that the Kane, 
Marsden and Ortiz's algorithm is a particular case of ours when we impose that the discrete 
lagrangian only depends on the time step ($h_k=t_{k+1}-t_k$).

Assume that
$
{L}_d(t_0, q_0,  t_1, q_1)=\bar{L}_d (q_0, q_1, t_1-t_0)=\bar{L}_d(q_0, q_1, h_0)
$
and, therefore, $D_1\bar{L}_d=D_2 L_d$, $D_2\bar{L}_d=D_4 L_d$, and  
\[
\frac{\partial \bar{L}_d}{\partial h}(q_0, q_1, t_1-t_0)=-
\frac{\partial L_d}{\partial t_0}(t_0, q_0,  t_1, q_1)=
\frac{\partial L_d}{\partial t_1}(t_0, q_0,  t_1, q_1)
\]
Then, equations (\ref{asd}) and (\ref{asd3}) can be rewritten as
\begin{eqnarray}
 h_kD_1\bar{L}_d(q_k, q_{k+1}, h_k)+h_{k-1}D_2\bar{L}_d(q_{k-1}, q_{k}, h_{k-1})&=&0\label{asd2}
\\
E_d(q_k, q_{k+1}, h_k)- E_d (q_{k-1}, q_k, h_{k-1})&=&0\label{asd4}
\end{eqnarray}
which are exactly the equations proposed in \cite{Kane}.

Observe that in this case the discrete lagrangian $L_d$ is invariant by time translations,
and the associated conservation law is precisely the energy.

\section{Concluding remarks}

This natural variational interpretation of the results of \cite{Kane} was 
independently obtained by Chen, Gou, Wu (see \cite{CGW}). 
They use Lee's approach and also derive in a variational way 
the symplectic-energy-momentum integrator for conservative lagrangian systems. 

The authors wish to thank an anonymous referee for valuable and 
useful comments and for pointing out to us reference \cite{CGW}.

\section*{Acknowledgments}
This work has been partially supported by grants DGICYT (Spain), PB97-1257
and PGC2000-2191-E.

\end{document}